\documentclass[3p,times,procedia]{elsarticle}

\UseRawInputEncoding
\usepackage{float}
\usepackage{amsmath,amssymb}
\usepackage{bm}
\usepackage{slashed}
\usepackage{braket}
\usepackage{epsfig}
\usepackage{graphicx}
\usepackage{hyperref}
\usepackage{xcolor}
\hypersetup{
    colorlinks=true,
    linkcolor=blue,
    filecolor=magenta,      
    urlcolor=blue,
    citecolor=blue
}
\newcommand{\comment}[1]{}

\begin{document}
\title{Thermalization of gluons in spatially homogeneous systems}
\author{Sergio Barrera Cabodevila}
\author{Carlos A. Salgado}
\author{Bin Wu}
\address{Instituto Galego de F\'isica de Altas Enerx\'ias IGFAE, Universidade de Santiago de Compostela, E-15782 Galicia-Spain}
\begin{abstract}
We investigate thermalization of gluons in spatially homogeneous systems using the Boltzmann equation in diffusion approximation. A complete picture on thermalization is obtained for both initially under- and over-populated systems. In an initially under-populated system, we find that its soft sector undergoes three stages: overheating, cooling/overcooling and reheating before full thermalization is achieved. In an initially over-populated system, we find that its soft sector only undergoes two stages towards full thermalization: overheating and cooling. The cooling stage is consistently driven by momentum broadening due to multiple elastic collisions, manifest as a non-thermal scaling solution.
\end{abstract}

\maketitle
\section{Introduction} 
One of the most salient features of heavy-ion collisions is the production of a dense system of gluons at very early stages of the collisions~\cite{Jalilian-Marian:1996mkd, Kovchegov:1998bi, Mueller:1999fp}. How these gluons evolve into a droplet of quark-gluon plasma poses one of the most challenging theoretical issues in heavy-ion physics~\cite{Schlichting:2019abc, Berges:2020fwq}. In the weak-coupling limit, such a dense system of gluons is revealed to establish thermal equilibrium in a "bottom-up" fashion ~\cite{Baier:2000sb}. The thermalization pattern in longitudinally boost-invariant systems has been subsequently explored in greater details using weak-coupling tools including classical statistical field theory~\cite{Berges:2013eia, Epelbaum:2013ekf} and effective kinetic theory (EKT)~\cite{Arnold:2002zm} in refs.~\cite{Kurkela:2015qoa, Kurkela:2018xxd, Kurkela:2018oqw, Du:2020dvp}. Classical statistical approximation is known to break down at later times~\cite{Mueller:2002gd, Berges:2013lsa, Epelbaum:2014yja,Epelbaum:2014mfa} and the EKT is so far the only weak-coupling tool that has been used to quantitatively study all the three stages of the bottom-up thermalization.

The Boltzmann equation in diffusion approximation (BEDA)~\cite{Mueller:1999pi, Baier:2000sb, Hong:2010at, Blaizot:2013lga} has been used to study certain aspects of thermalization.  Without number-change processes, the onset of a Bose-Einstein condensate (BEC) of gluons is observed to occur within a finite time in the soft sector if the system is initially over-populated~\cite{Blaizot:2013lga}. In this case, the BEDA admits unique solutions composed of a phase-space distribution for $p>0$ and a BEC: a macroscopic accumulation of gluons at $p=0$~\cite{Blaizot:2014jna}, which agrees, at least qualitatively, with the full Boltzmann equation including only elastic collisions~\cite{Xu:2014ega}. Another quantitative study using the diffusion approximation is to include the so-called Bethe-Heitler (BH) rate in the $2\leftrightarrow3$ process~\cite{Huang:2013lia, Blaizot:2016iir}, which  is shown to strongly hinder the formation of the BEC~\cite{Blaizot:2016iir}. In ref.~\cite{Baier:2000sb}, the BEDA including both the elastic scattering and the inelastic scattering with the QCD analogue of  
the Landau-Pomeranchuk-Migdal (LPM) effect~\cite{Baier:1996kr, Baier:1998kq}  was used to justify the parametric estimates of the bottom-up thermalization. This equation, bearing several main physical ingredients in common with the EKT, has not yet been solved numerically or used to quantitatively study thermalization.

Compared to the EKT, the BEDA, with its collision kernels greatly simplified and its solution less numerical demanding, could be an alternative, practical tool to study thermalization/hydrodynamization in 2+1D or 3+1D systems and to describe collective flow phenomenologically in small systems. We leave such studies for future publications. The main purpose of this work is to show that the diffusion approximation is a well justified simplification which does not incur any unwanted, qualitatively different features of thermalization compared to the EKT. We elaborate this point by a detailed comparison of qualitative features of thermalization in spatially homogeneous systems of gluons predicted by the EKT and the BEDA based on parametric estimates. Such parametric estimates are verified quantitatively by numerically solving the BEDA including both the elastic scattering~\cite{Mueller:1999pi, Hong:2010at, Blaizot:2013lga} and the inelastic scattering with the LPM effect~\cite{Baier:2000sb} for the first time.

As a summary of our results detailed below, we find the following complete picture on thermalization of gluons in spatially homogeneous systems in terms of both soft and hard sectors described by the BEDA:

An initially under-populated system with the initial occupation number $f_0\ll 1$ thermalizes through three stages. The system first goes through an {\it overheating} stage in its soft sector. Starting from the initial time $t=0$, soft gluons are rapidly generated by radiation off hard gluons (whose momenta are characterised by $Q$).  They fill a thermal distribution with the highest temperature $T_*\sim\,Q$ for $p\lesssim p_*$. Here, $p_*$ is a soft momentum scale mainly determined by the inelastic scattering. The hard sector predominantly determines all the relevant macroscopic properties of the system including $T_*$ until $t\sim \alpha_s^{-2} f_0 Q^{-1}$. Then, the soft sector undergoes a {\it cooling} stage.  At $t\sim \alpha_s^{-2} f_0 Q^{-1}$, soft gluons start to contribute predominantly to, and hence enhance, the color screening effect. This implies that the mean free path of gluons, denoted by $\lambda$, is increased and the temperature of the soft sector $T_*\propto 1/(\alpha_s\lambda)$ is decreased accordingly. During this stage, the number of soft gluons is still much lower than that of hard ones until $t\sim \alpha_s^{-2} f_0^{-\frac{1}{3}}Q^{-1}$. At this momentum, the soft sector is in an overcooled state with $T_*\ll T_{eq}$, the final thermal equilibrium temperature, and all the relevant macroscopic quantities except the energy density start to receive dominant contributions from the soft sector. The only major role of hard gluons is then to inject energy into the soft sector and the system undergoes a {\it reheating} stage with an increasing $T_*$. Full thermalization is achieved at $t\sim \alpha_s^{-2} f_0^{-\frac{3}{8}} Q^{-1}$ when hard gluons lose all their energy to, and heat up again, the soft thermal bath. Such qualitative features, in agreement with the EKT~\cite{Kurkela:2014tea, Schlichting:2019abc, Du:2020dvp}, are verified below by quantitative studies using the BEDA.

An initially over-populated system with $f_0\gg 1$ thermalizes through two stages. As in the under-populated case, the system first goes through an {\it overheating} stage in its soft sector with temperature $T_*\sim f_0 Q$, which prevents the formation of a BEC. $T_*$ is predominantly determined by the hard sector and kept as a constant parametrically until $t\sim (\alpha_s f_0)^{-2}Q^{-1}$. Around this time, all gluons typically pick up some momentum broadening of $O(Q)$ due to multiple elastic scatterings. Afterwards, the typical momentum of gluons is consistently given by multiple scatterings and the number density of gluons has to decrease accordingly as mandated by energy conservation. As a result, $T_*$ cools down and  all the quantities in the system evolve according to a set of scaling laws, manifest as a universal scaling solution previously discovered in both classical statistical field simulations~\cite{Berges:2012ev, Schlichting:2012es, Kurkela:2012hp} and EKT~\cite{AbraaoYork:2014hbk, Kurkela:2014tea, Du:2020dvp}. The above argument is also true and, hence, the scaling solution also holds if one includes only the elastic scattering with a BEC in the BEDA as prescribed in ref.~\cite{Blaizot:2014jna} (see ref.~\cite{Blaizot:2011xf} for an alternative argument) or replaces the LPM rate by the BH rate in the inelastic kernel.

\section{The QCD Boltzmann equation}
The Boltzmann equation at leading order takes the form~\cite{Baier:2000sb, Arnold:2002zm}
\begin{align}
\label{eq:boltz1}
\left(\partial_t + {\bm v}\cdot \nabla_{\bm x}\right) f =&C_{el}[f] + C_{inel}[f],
\end{align}
where the $2\leftrightarrow2$ scattering and the $1\leftrightarrow2$ processes are described by $C_{el}$ and $C_{inel}$ respectively. There are two main quantities that determine qualitative features of thermalzation: the Debye mass $m_D$ and the jet quenching parameter $\hat{q}$~\cite{Baier:1996sk}:
\begin{align}\label{eq:mDqhatDef}
    m_D^2 = 8\pi \alpha_s N_c\int\frac{d^3 \bm{p}}{(2\pi)^3}\frac{f}{|\bm{p}|},\qquad
    \hat{q} = 8\pi N_c^2\alpha_s^2 \ln\frac{\langle p_t^2\rangle}{m_D^2}\int\!{d^3{\bm p}\over(2\pi)^3}f\left(1+f\right)\qquad\text{for pure gluon systems}
\end{align}
with $\langle p_t^2\rangle$ the average transverse momentum broadening.

The $2\leftrightarrow2$ processes in QCD are dominated by small-angle scatterings. In BEDA, the elastic kernel is determined solely by small-angle scatterings, which explicitly depends on $\hat{q}$ and $m_D$~\cite{Mueller:1999pi, Hong:2010at, Blaizot:2013lga}:
\begin{align}\label{eq:Cel}
C_{el}[f]=\frac{1}{4}\hat{q}(t)\nabla_{{\bm p}} \cdot\left[  \nabla_{{\bm p}}f  + \frac{{\bm v}}{T_*(t)}f(1+f)\right],
\end{align}
with the effective temperature
\begin{align}\label{eq:TstarDef}
    T_*(t)\equiv\frac{\hat{q}}{2\alpha_s N_C  m_D^2\ln\frac{\langle p_t^2\rangle}{m_D^2}}.
\end{align}
In contrast, the elastic kernel in EKT describes also wide-angle scatterings, which are, however, irrelevant to the investigation of qualitative features of thermalization below. Since small-angle scatterings are only realized if the Debye mass is parametrically smaller than the typical momentum of the particles, $\alpha_s f_0$ is assumed not to be larger than unity in the following discussions.

The inelastic kernel takes a generic form~\cite{Baier:2000sb, Arnold:2002zm}
\begin{align}
   C_{inel}[f_{\bm{p}}] = \int\,d^3\bm{p}'\int\limits_0^1\!dx\, {d^2 I(\bm{p}') \over dx\, dt}\bigg\{&
   \delta(\bm{p}-x\bm{p}')
   \biggl[ f_{\bm{p}'}(1+f_{x\bm p'}) \bigg(1+f_{(1-x)\bm{p}'}\bigg) -
   f_{x\bm p'}f_{(1-x){\bm p'}}\left(1{+}f_{{\bm p'}}\right)\biggr] \nonumber\\
   &-{1\over2}\delta(\bm{p}-\bm{p}') \biggl[f_{\bm p}(1{+}f_{x{\bm p}})\left(1{+}f_{(1{-}x){\bm p}}\right)
     -f_{x{\bm p}}f_{(1{-}x){\bm p}}(1{+}f_{\bm p})\biggr]
   \biggr\}, \label{eq:Cinel}
\end{align}
where a short-hand notation $f_{\bm{p}}$ is used to show explicitly the dependence of $f$ on momentum $\bm{p}$ and ${d^2 I(\bm{p}')/ dx\, dt}$ is the rate for one gluon of momentum $\bm{p}'$ to split into two gluons carrying momentum fraction $x$ and ($1-x$) respectively. The $1\leftrightarrow2$ scattering due to the LPM effect is the most efficient mechanism to drive the system towards full thermalization~\cite{Baier:2000sb, Kurkela:2014tea}. Accordingly, the splitting rate due to the LPM effect~\cite{Baier:1996kr, Baier:1998kq}
\begin{equation}\label{eq:ILPM}
   \frac{d^2I(p)}{dxdt} = \frac{\alpha_s N_c}{\pi}{(1-x+x^2)^{5/2}\over(x-x^2)^{3/2}}\sqrt{\frac{\hat{q}}{p}}
\end{equation}
is used for quantitative studies in BEDA~\cite{Baier:2000sb}. It is the same as that in EKT in the deep LPM regime~\cite{Arnold:2008zu} while in EKT the splitting rate switches smoothly to the BH rate for~\cite{Baier:1996kr}
\begin{align}
x p \lesssim p_{LPM}\equiv \hat{q}\lambda^2 = \frac{m_D^4}{\hat{q}}
\end{align}
with the mean free path $\lambda = m_D^2/\hat{q}$.
  
\section{Rapid thermalization in the soft sector}
The soft sector maintains a thermal distribution with temperature $T_*$ during the entire thermalization process. Let us first focus on very early times. In the inelastic kernel with the LPM rate, the leading-order term in $1/p$ at small $p$ is given by
\begin{align}
   C_{inel}[f_{\bm{p}}]&\approx \frac{\alpha_s N_c}{\pi}\sqrt{\frac{\hat{q}}{p}}\int\limits_0^1\frac{dx}{x^4}
   \biggl[ p f_{\bm{p}}f'_{\frac{\bm{p}}{x}}+f_{\frac{\bm{p}}{x}}(1+f_{\frac{\bm{p}}{x}})-p f_{\frac{\bm{p}}{x}}f'_{\frac{\bm{p}}{x}}\biggr]
   \biggr\}\approx\frac{\alpha_s N_c}{\pi}I_a\sqrt{\frac{\hat{q}}{p}}\frac{1}{p^3}\bigg(1-\frac{pf_{\bm{p}}}{T_*}\bigg)
\end{align}
with $I_a \equiv\hat{q}/\bigg(\frac{4\alpha_s^2 N_c^2}{\pi} \ln\frac{\langle p_t^2\rangle}{m_D^2}\bigg)$ and the derivative with respect to $p$ denoted by primes. Here and below we only consider the cases with $f$ dependent of $p=|\bm{p}|$. Neglecting the elastic kernel and the time-dependence of $\hat{q}$ and $m_D$, the BEDA admits an approximate solution
\begin{align}\label{eq:pfapprox}
p f_{\rm p}(t) &\approx T_*\bigg[1-\bigg(1-\frac{p f_p(t=0)}{T_*}\bigg)e^{-\frac{1}{2}\left(\frac{p_*}{p}\right)^{\frac{5}{2}}}\bigg] \approx
\left\{\begin{array}{ll}
T_*     &  \text{for $p\lesssim p_*$}\\
pf_p(t=0) + \frac{T_*}{2}\left(\frac{p_*}{p}\right)^{\frac{5}{2}} & \text{for $p\gtrsim p_*$} 
\end{array}
\right..
\end{align}
Since $T_*\gg p_*$, it reveals that soft gluons fill a thermal distribution with temperature $T_*$ up to a new soft momentum scale
\begin{align}\label{eq:pstar}
p_* \equiv (\hat{q} m_D^4 t^2)^{\frac{1}{5}}
\end{align}
while the sector with $p\gtrsim p_*$ is populated predominantly by single gluon radiation due to the LPM effect. In contrast, the soft sector at very early times is mainly generated by the BH rate in EKT. It was shown that the soft sector generated by the BH rate thermalizes up to momentum $\propto(\hat{q}t)^{\frac{1}{2}}$ with temperature also given by $T_*$~\cite{Blaizot:2016iir}. Therefore, in both BEDA and EKT, one has $f\to{T_*}/{p}$ as $p\to0$. Since the number density of the BEC transiently formed per unit time is proportional to ($\lim\limits_{p\to0}pf-T_*$)~\cite{Blaizot:2014jna}, one can conclude that irrespective of the initial occupation number there is no transient formation of the BEC in both BEDA and EKT\footnote{The full LPM spectrum is less divergent at small $p$~\cite{Salgado:2003gb}, which, however, will not change our conclusion as long as soft radiation dominates.
}.

The difference in the typical momentum of soft gluons predicted by the BEDA and the EKT persists until
\begin{align}\label{eq:tLPM}
    t\sim\,t_{LPM}\equiv\frac{m_D^8}{\hat{q}^3},\qquad\,\text{with~}p_{*}\sim p_{LPM}= \frac{m_D^4}{\hat{q}}
\end{align}
when $p_*\sim(\hat{q}t)^\frac{1}{2}$. Below, we consider a generic, spatially homogeneous system of gluons with an initial phase-space distribution characterised by $f_0$ and the initial momentum by $Q$ to investigate whether such a difference results in qualitatively different features of thermalization.

\section{Initially under-populated systems}
\subsection{Qualitative features}
In the limit $f_0\ll 1$, according to eqs.~(\ref{eq:mDqhatDef}) and (\ref{eq:TstarDef}) the system initially has
\begin{align}\label{eq:quantities0_under}
n\sim f_0 Q^3,~~m_{D}^2\sim \alpha_s f_0 Q^2,~~\hat{q}\sim \alpha_s^2 f_0 Q^3,~~T_*\sim Q
\end{align}
with $n$ the gluon number density. By using energy conservation with the energy density $\epsilon\sim f_0 Q^4$, one can predict their final thermal equilibrium values
\begin{align}\label{eq:quantitieseq}
T_{eq}\sim \epsilon^{\frac{1}{4}}\sim f_0^{\frac{1}{4}}Q,~~n_{eq}\sim f_0^{\frac{3}{4}}Q^3,~~m_{Deq}^2\sim \alpha_s f_0^{\frac{1}{2}}Q^2,~~\hat{q}_{eq}\sim \alpha_s^2 f_0^{\frac{3}{4}}Q^3.
\end{align}
With these quantities evolving towards their thermal equilibrium values the system goes through three stages:

{\it 1. Soft gluon radiation and overheating: $0\ll\,Qt\ll \alpha_s^{-2} f_0$.} Assuming the Debye mass and $\hat{q}$ given by their initial parametric forms in eq.~(\ref{eq:quantities0_under}), one has from eq.~(\ref{eq:tLPM}) 
\begin{align}
t_{LPM}\sim \alpha_s^{-2}f_0 Q^{-1},\qquad\,p_{LPM}\sim f_0 Q.
\end{align}
Before $t\sim t_{LPM}$, the typical momenta of soft gluons generated via radiation in BEDA and EKT are different as discussed previously:
\begin{align}
p_s\sim\left\{
\begin{array}{ll}
(\hat{q}t)^{\frac{1}{2}}\sim\alpha_s f_0^{\frac{1}{2}} (Q t)^{\frac{1}{2}}Q&\text{for the BH rate in EKT}\\
p_* = (\hat{q} m_D^4 t^2)^{\frac{1}{5}}\sim \alpha_s^{\frac{4}{5}}f_0^{\frac{3}{5}} (Q t)^{\frac{2}{5}}Q &\text{for the LPM rate in BEDA}
\end{array}
\right.
.
\end{align}
In BEDA, $p_s$ is given by $p_*$ because momentum broadening from multiple elastic scatterings $\sim(\hat{q}t)^{\frac{1}{2}}$ is parametrically smaller, meaning the elastic scattering modifies the approximate solution in eq.~(\ref{eq:pfapprox}) only at $p\ll p_*$. Accordingly, the number density of soft gluons can be estimated by the BH and LPM splitting rates respectively in EKT and BEDA~\cite{Baier:1996kr}
\begin{align}\label{eq:nsI}
\frac{n_s}{n_h}\sim
 x\frac{dI(Q)}{dxdt}t\sim\left\{
\begin{array}{ll}
\frac{\alpha_s}{\lambda} t\sim\alpha_s\frac{\hat{q}}{m_D^2} t\sim \alpha_s^2 (Qt)&\text{in EKT}\\
\alpha_s\sqrt{\frac{\hat{q} t^2}{p_*}}\sim \alpha_s^{\frac{8}{5}} f_0^{\frac{1}{5}} (Qt)^{\frac{4}{5}}&\text{in BEDA}
\end{array}
\right.,
\end{align}
which can be verified by using eq.~(\ref{eq:pfapprox}) in the case of BEDA. Here, the number density of hard gluons $n_h\sim\,n\sim f_0 Q^3$ and $xQ\sim\,p_s$. In both theories, one has
\begin{align}\label{eq:fsI}
f_s\sim\frac{n_s}{p_s^3}\sim\frac{T_*}{p_s}\gg 1,
\end{align}
which shows consistently thermalization in the soft sector with temperature $T_*$. Both $\hat{q}$ and $m_D^2$ receive corrections from soft gluons with
\begin{align}\label{eq:under_qmD2I}
\frac{\hat{q}_s}{\hat{q}_h}\sim \frac{ n_s f_s}{n_h}\sim\left\{
\begin{array}{ll}
\alpha_s f_0^{-\frac{1}{2}}(Q t)^{\frac{1}{2}}&\text{in EKT}\\
\alpha_s^{\frac{4}{5}} f_0^{-\frac{2}{5}} (Qt)^{\frac{2}{5}}&\text{in BEDA}
\end{array}
\right.
,\qquad
\frac{m_{Ds}^2}{m_{Dh}^2}\sim \frac{n_s Q}{n_h p_s}\sim\left\{
\begin{array}{ll}
\alpha_s f_0^{-\frac{1}{2}}(Q t)^{\frac{1}{2}}&\text{in EKT}\\
\alpha_s^{\frac{4}{5}}f_0^{-\frac{2}{5}} (Q t)^{\frac{2}{5}}&\text{in BEDA}
\end{array}
\right. 
.
\end{align}
Here and below the contributions from hard and soft gluons to macroscopic quantities are denoted by subscripts $h$ and $s$ respectively.
Since the above ratios are both parametrically small at $Q t\ll \alpha_s^{-2}f_0$, all the macroscopic quantities in the system are parametrically the same as their initial forms in eq. (\ref{eq:quantities0_under}), hence justifying the above analysis with $\hat{q}$ and $m_D^2$ kept fixed. Especially, the thermalized soft sector is overheated with $T_*\sim Q\gg\,T_{eq}$. This is true in both BEDA and EKT. 

{\it 2. The cooling and overcooling of soft gluons: $\alpha_s^{-2} f_0\ll\,Qt\ll \alpha_s^{-2} f_0^{-\frac{1}{3}}$.}
Starting from $Qt\sim \alpha_s^{-2} f_0$, soft radiation is typically given by the LPM effect both in BEDA and EKT and the subsequent parametric analysis in BEDA is identical to that in EKT~\cite{Schlichting:2019abc}. $p_*$ is not relevant any more and the typical momentum of soft gluons $p_s$ is pushed up significantly by multiple elastic scatterings, yielding
\begin{align} 
p_s\sim (\hat{q}t)^{\frac{1}{2}}\sim \alpha_s f_0^{\frac{1}{2}} (Q t)^{\frac{1}{2}}Q\gg p_*.
\end{align}
As a result, the number density and the phase-space distribution of soft gluons are given by
\begin{align}
\frac{n_s}{n_h}\sim \alpha_s\sqrt{\frac{\hat{q} t^2}{p_s}}\sim \alpha_s^{\frac{3}{2}}f_0^{\frac{1}{4}}(Qt)^{\frac{3}{4}}\ll 1,\qquad
f_s\sim\frac{n_s}{p_s^3}\sim \frac{f_0^{-\frac{1}{4}}}{\alpha_s^{\frac{3}{2}}(Qt)^{\frac{3}{4}}}\gg 1
\end{align}
for $Qt\ll \alpha_s^{-2} f_0^{-\frac{1}{3}}$. $\hat{q}$ receives parametrically equal contributions from soft and hard gluons while the Debye mass and $T_*$ receive predominant contributions from soft gluons:
\begin{align}\label{eq:mD2II}
\hat{q}\sim \alpha_s^2 f_0 Q^3,\qquad\,m_D^2 \sim \frac{\alpha_s n_s}{p_s} \sim \alpha_s^{\frac{3}{2}}f_0^{\frac{3}{4}} (Qt)^{\frac{1}{4}} Q^2,\qquad
T_*\sim \frac{\hat{q}}{\alpha_s m_D^2} \sim \alpha_s^{-\frac{1}{2}}f_0^{\frac{1}{4}}(Qt)^{-\frac{1}{4}}Q.
\end{align}
These parametric estimates are consistent with the fact that the soft sector maintains a thermal distribution with $f_s\sim n_s/p_s^3\sim T^*/p_s$ and starts to become overcooled 
at $Qt\sim \alpha_s^{-2}$ when $T_*\sim T_{eq}\sim f_0^{\frac{1}{4}}Q$.

At $Qt\ll \alpha_s^{-2} f_0^{-\frac{1}{3}}$, the typical energy carried by each soft gluon $\sim\,p_s$ is parametrically larger than the typical energy loss of a hard gluon~\cite{Baier:2000sb, Baier:2001yt}
\begin{align}
p_{\text{br}}\sim \alpha_s^2 \hat{q}t^2\sim \alpha_s^4 f_0 (Qt)^2 Q
\end{align}
as a result of democratic branching~\cite{Blaizot:2013hx}. Since $p_{\text{br}}\ll p_s$ and $T_*$, the conditions for democratic branching are not satisfied~\cite{Iancu:2015uja} and, therefore, it is not the dominant mechanism for radiative energy loss of hard gluons at this stage. Only at the end of this stage around $Q t\sim \alpha_s^{-2} f_0^{-\frac{1}{3}}$, the number of soft gluons becomes comparable with that of hard ones and the soft sector is overcooled to the lowest temperature $T_*\sim f_0^{\frac{1}{3}}Q$. Meanwhile, the relaxation time of soft gluons becomes comparable with $t$ and $p_{\text{br}}\sim p_s\sim T_*$. This implies that soft gluons are poised to achieve thermal equilibration among themselves and democratic branching is about to dominate the energy loss of hard gluons.

{\it 3. Reheating of soft gluons and mini-jet quenching: $\alpha_s^{-2} f_0^{\frac{1}{3}}\ll\,Qt\ll \alpha_s^{-2} f_0^{-\frac{3}{8}}$.}
Starting from $Q t\sim \alpha_s^{-2} f_0^{-\frac{1}{3}}$, the number of soft gluons becomes larger than that of hard ones. As a result, both the jet quenching parameter and the Debye mass
start to receive dominant contributions from soft gluons. Moreover, there is enough time for soft gluons to establish thermal equilibration among themselves with the energy density $\epsilon_s\sim\,T_*^4$. On the other hand, $\epsilon_s$ is predominantly determined by the typical energy loss of hard gluons with
$
\epsilon_s\sim\,p_{\text{br}} n_h\sim \alpha_s^4 T_*^3 t^2 f_0 Q^3.
$
Equating the above two estimates of $\epsilon
_s$ yields~\cite{Kurkela:2014tea}
\begin{align}\label{eq:quantities3_under}
T_*\sim \alpha_s^4 f_0 (Qt)^2 Q,\qquad
m_D^2\sim \alpha_s T_*^2\sim \alpha_s^9 f_0^2 (Qt)^4 Q^2,\qquad \hat{q}\sim \alpha_s^2 T_*^3\sim \alpha^{14} f_0^3(Qt)^6 Q^3.
\end{align}
Thermalization is completed when hard gluons lose all their energies at $Qt\sim \alpha_s^{-2} f_0^{-\frac{3}{8}}$.

\subsection{Quantitative results}
\begin{figure}
\begin{center}
\includegraphics[width=0.49\textwidth]{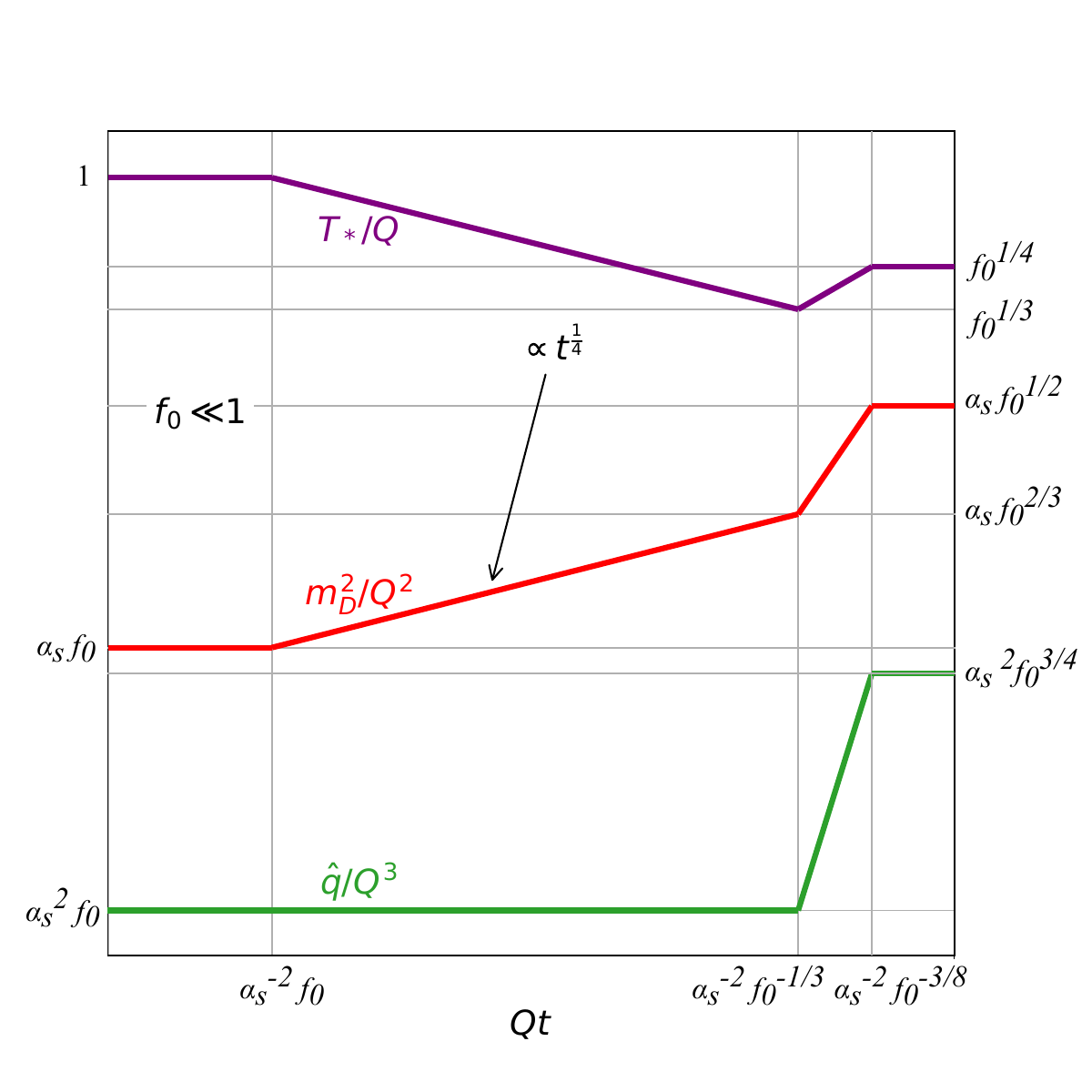}
\includegraphics[width=0.49\textwidth]{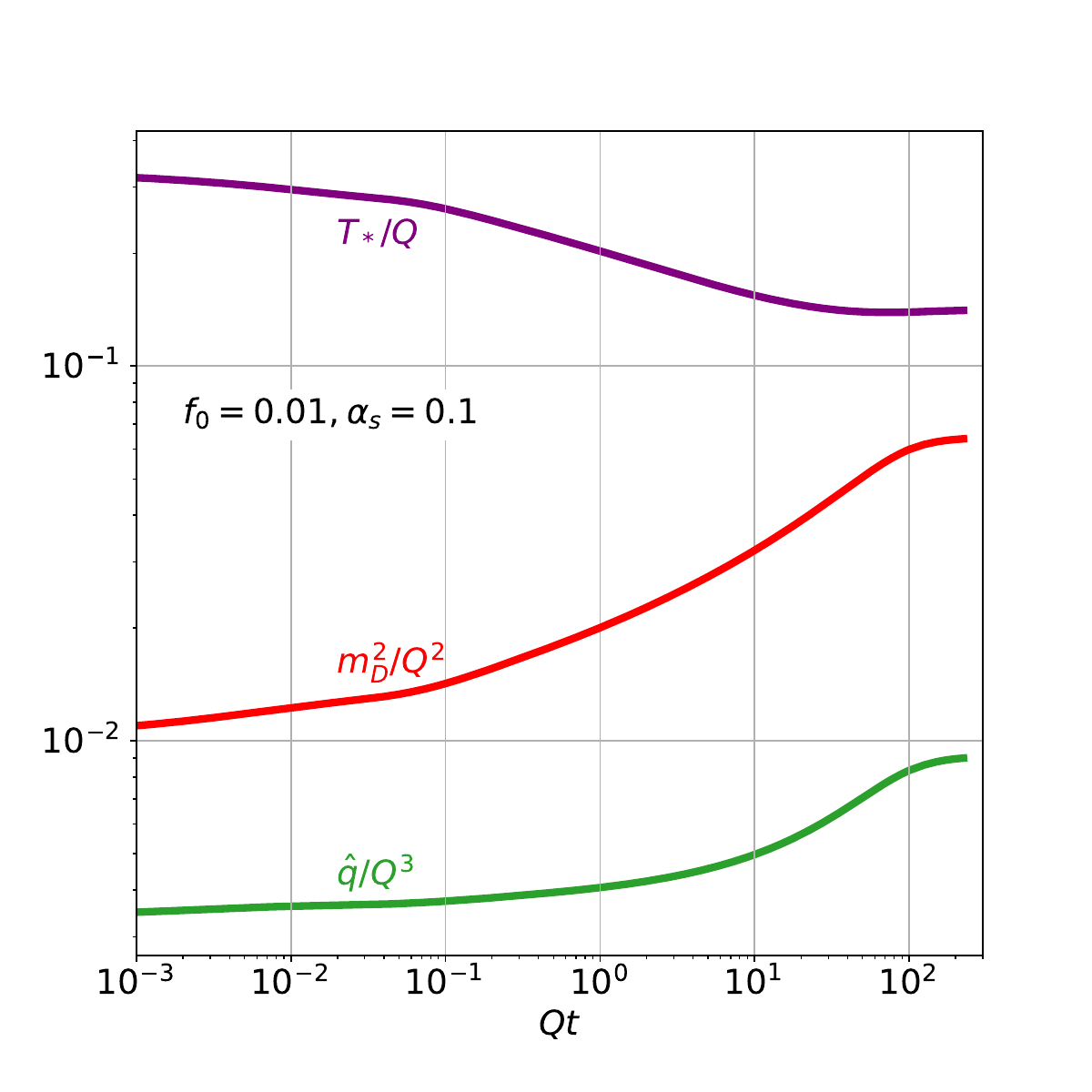}
\caption{The time evolution of $T_*$, $m_D^2$ and $\hat{q}$ in initially under-populated systems. Left panel: the distinct three stages in the limit $f_0\ll1$, respectively given by eqs.~(\ref{eq:quantities0_under}), (\ref{eq:mD2II}) and (\ref{eq:quantities3_under}). Right panel: numerical results for $f_0=0.01, \alpha_s=0.1$.
\label{fig:underpopulated}}
\end{center}
\end{figure}

\begin{figure}
\begin{center}
\includegraphics[width=0.49\textwidth]{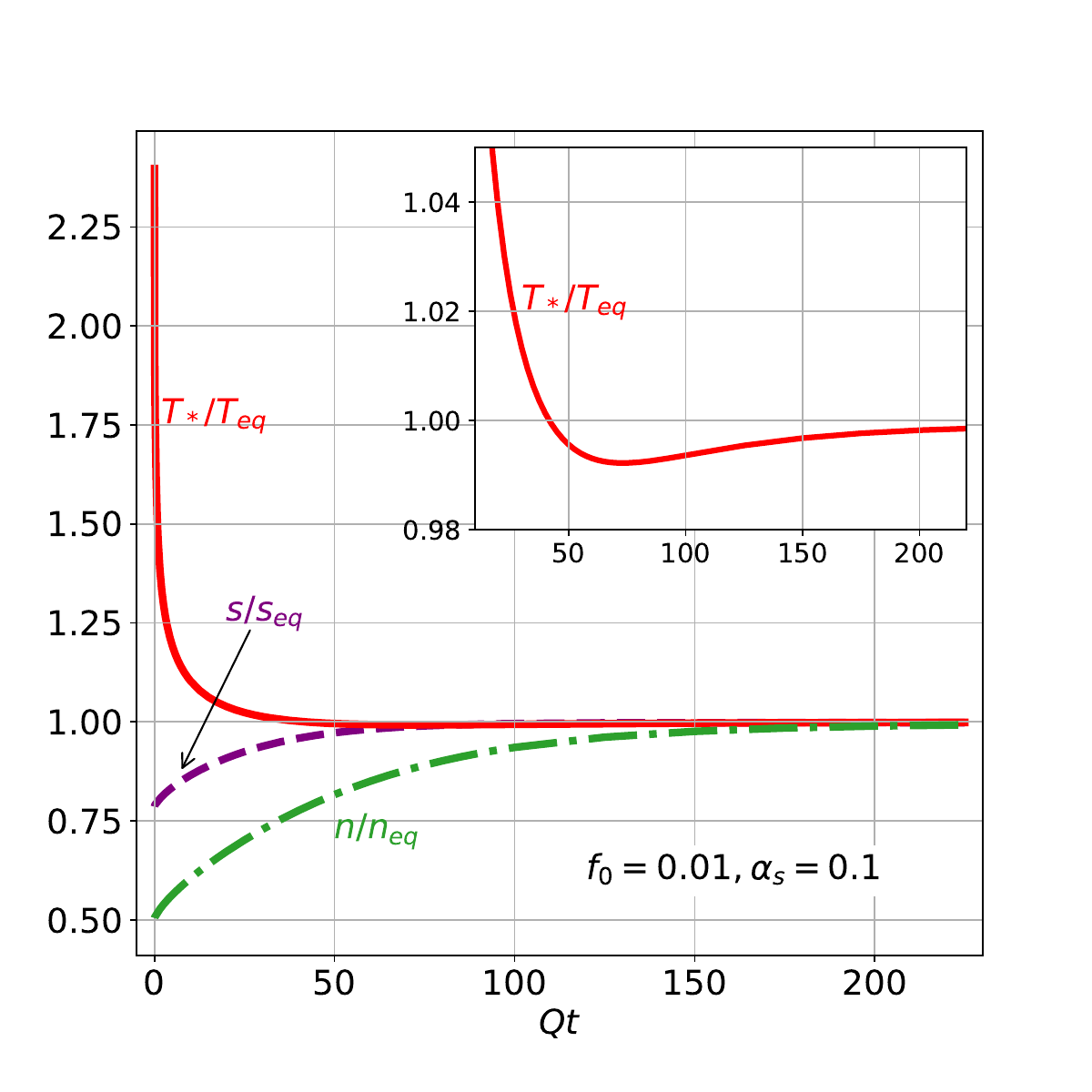}
\includegraphics[width=0.49\textwidth]{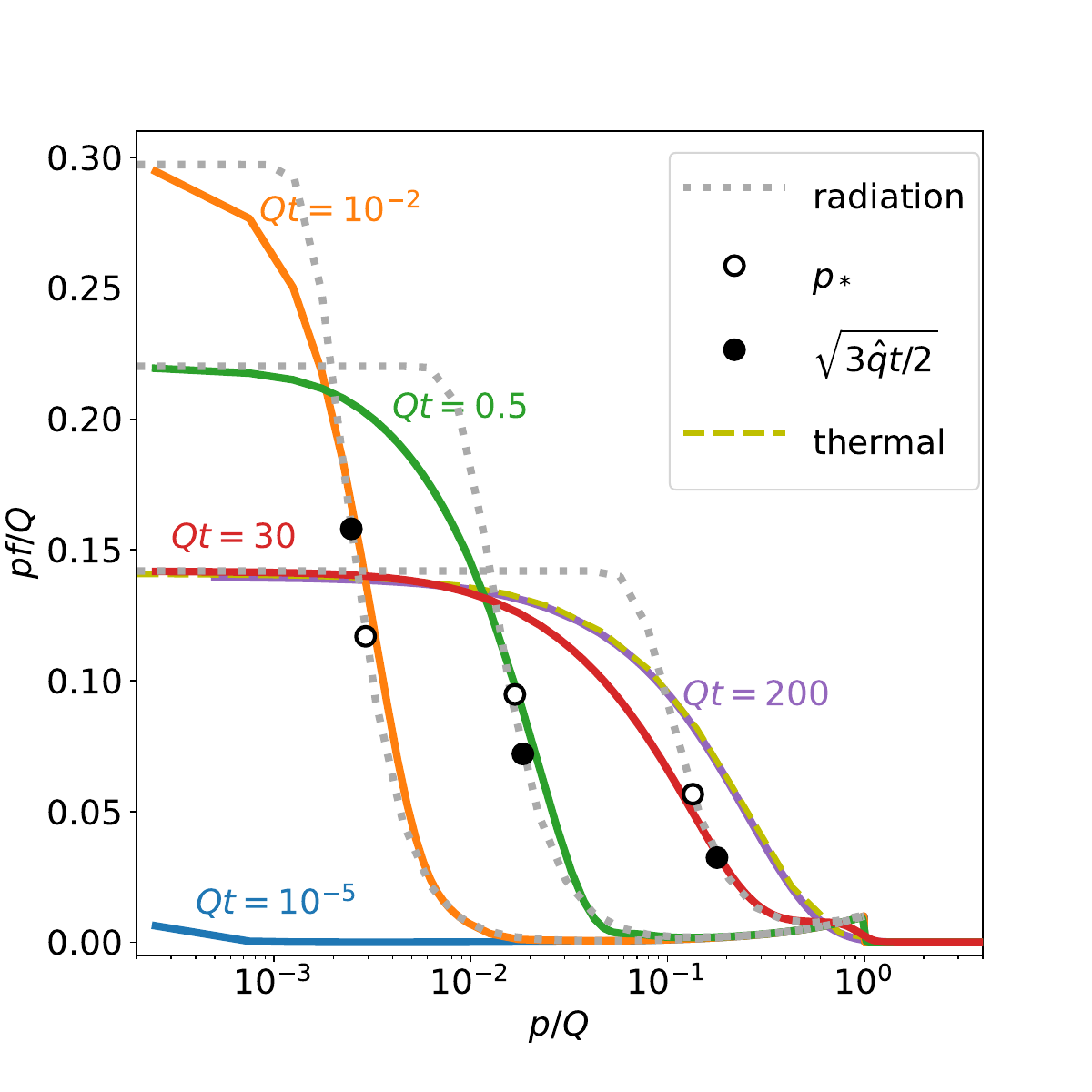}
\caption{Features of thermalization  for $f_0=0.01$ and $\alpha_s=0.1$. Left panel: the temperature $T_*$, the entropy density $s$ and the number density $n$ as a function of $t$. Right panel: the phase-space distribution $pf$ at different times. The dashed curves (radiation) are given by the approximate solution purely from radiation in eq.~(\ref{eq:pfapprox}) with locations of $p_*$ (open) and $\sqrt{3\hat{q}t/2}$ (closed) indicated by dots on them. $pf$ at $Qt=200$ fits very well with the thermal distribution (thermal) with temperature $T_{eq}$ predicted by energy conservation.
\label{fig:f0_0p01}}
\end{center}
\end{figure}

For quantitative studies, we solve numerically the BEDA with the initial distribution given by
$
f(t_0) = {f_0}\theta(Q-|\vec{p}|).
$
Figure \ref{fig:underpopulated} compares the time evolution of $\hat{q}$, $m_D^2$ and $T_*$ for $f_0=0.01$ and $\alpha_s=0.1$ (right panel) to those given by the above parametric estimates (left panel). For such an intermediate value of $f_0$, the system still undergoes, although not as pronounced as our parametric estimates, the three stages manifested in the macroscopic quantities discussed above
. Combined with the left panel of fig.~\ref{fig:f0_0p01}, the temperature curve confirms more convincingly the three-stage thermalization: overheating, cooling with a small overcooling effect\footnote{
  We have verified numerically that the overcooling effect becomes more pronounced for smaller values of $f_0$. One can also estimate parametrically that the system suffers the maximum temperature overshoot with $T_*/T_{eq}\sim f_0^{\frac{1}{12}}$ at $Qt\sim \alpha_s^{-2}f_0^{-\frac{1}{3}}$.
} and reheating. The three stages are also manifest in the time evolution of $pf$ as shown in the right panel of fig.~\ref{fig:f0_0p01}: $\lim\limits_{p\to0}pf=T_*$ goes through all the three stages. All the quantities including $T_*$, the entropy density $s$, the number density $n$ and $pf$ stabilize around their corresponding thermal equilibrium values at $Qt\gtrsim 200$.

We also check quantitatively the role played by $p_*$ and momentum broadening $\sqrt{\langle\,p^2\rangle}\equiv\sqrt{3\hat{q}t/2}$. We find that $\sqrt{\langle\,p^2\rangle}$  becomes larger than $p_*$ at $Q t \approx 0.1$ for $f_0=0.01$ and $\alpha_s=0.1$. As shown in the right panel of fig.~\ref{fig:f0_0p01}, up to this time (and a bit later) the approximate solution with the scaling behavior $f\propto p^{-\frac{7}{2}}$ in eq.~(\ref{eq:pfapprox}) agrees with our simulations very well for $p\gtrsim p_*$. For $Qt\gg 0.1$, the approximate solution fits our simulations only for $p\gtrsim\sqrt{\langle\,p^2\rangle}$ while $pf$ in our simulations is strongly flatten at smaller $p$. This justifies quantitatively our parametric estimates for $p_s\sim(\hat{q}t)^{\frac{1}{2}}$ instead of $p_*$ in stage 2.

\section{Initially over-populated systems}
\subsection{Qualitative features}
In the limit $f_0\gg 1$ with $\alpha_s f_0\lesssim1$, the system evolves from an initial distribution of gluons with
\begin{align}\label{eq:macsOverI}
n\sim f_0 Q^3,~~m_{D}^2\sim \alpha_s f_0 Q^2,~~\hat{q}\sim \alpha_s^2 f_0^2 Q^3, ~~ T_*\sim f_0 Q
\end{align}
towards a full thermal distribution with their thermal equilibrium values given by the same expressions as the under-populated case in eq.~(\ref{eq:quantitieseq}). In this case, the system thermalizes through two stages:

{\it 1. Soft gluon radiation and overheating: $0\ll\,Qt\ll (\alpha_s f_0)^{-2}$.} From eq.~(\ref{eq:macsOverI}), one has
\begin{align}
t_{LPM}\sim (\alpha_s f_0)^{-2}Q^{-1},\qquad p_{LPM}\sim Q.
\end{align}
Therefore, one needs to distinguish between the BEDA and the EKT during this stage. 
In both theories, a soft sector is built up rapidly to fill a thermal distribution for $p\lesssim p_s$ with
\begin{align}
p_s\sim\left\{
\begin{array}{ll}
(\hat{q}t)^{\frac{1}{2}}\sim\alpha_s f_0 (Qt)^{\frac{1}{2}}Q&\text{for the BH rate in EKT}\\
p_* = (\hat{q} m_D^4 t^2)^{\frac{1}{5}}\sim (\alpha_s\,f_0)^{\frac{4}{5}}(Qt)^{\frac{2}{5}} Q &\text{for the LPM rate in BEDA}
\end{array}
\right.
.
\end{align}
And the system, albeit over-populated, witnesses an increase of the number of gluons~\cite{Huang:2013lia, Blaizot:2016iir} with the number of soft gluons estimated to be
\begin{align}
\frac{n_s}{n_h}\sim
 x\frac{dI(Q)}{dxdt}f_0 t\sim
\left\{
\begin{array}{ll}
\frac{\alpha_s}{\lambda} f_0 t\sim\alpha_s\frac{\hat{q}}{m_D^2} f_0 t\sim (\alpha_s f_0)^2 (Qt)&\text{in EKT}\\
\alpha_s\sqrt{\frac{\hat{q} t^2}{p_*}} f_0\sim (\alpha_s f_0)^{\frac{8}{5}} (Qt)^{\frac{4}{5}} &\text{in BEDA}
\end{array}
\right.
,
\end{align}
which is consistent with a thermal distribution:
\begin{align}
f_s\sim \frac{n_s}{p_s^3} \sim\frac{T_*}{p_s}.
\end{align}
Before $Qt\sim(\alpha_s f_0)^{-2}$, the contributions from soft gluons to $n$, $m_D^2$ and $\hat{q}$ are all parametrically small. Therefore, all the relevant macroscopic quantities in both BEDA and EKT are parametrically given by eq.~(\ref{eq:macsOverI}) throughout this stage.

{\it 2. Momentum broadening and cooling: $(\alpha_s f_0)^{-2}\ll\,Qt\ll \alpha_s^{-\frac{7}{4}}(\alpha_s f_0)^{-\frac{1}{4}}$.}
Starting from $Qt\sim (\alpha_s f_0)^{-2}$, the BEDA and the EKT admit the same parametric estimates. The typical momentum of gluons is mainly determined by multiple elastic scatterings and the total energy is conserved, yielding
\begin{align}
    p^2 \sim \hat{q} t \sim\alpha_s^2 n^2 t/p^3,\qquad
    \epsilon\sim\,f_0 Q^4 \sim\,n p.
\end{align}
From these two equations, one can solve self-consistently
\begin{align}\label{eq:macsOverII}
&n\sim \frac{(\alpha_s f_0)^{\frac{5}{7}}}{\alpha_s}\frac{Q^3}{ (Qt)^\frac{1}{7}}, \qquad p\sim (\alpha_s f_0)^{\frac{2}{7}}(Qt)^{\frac{1}{7}} Q,\qquad\,f\sim \frac{(\alpha_s f_0)^{-\frac{1}{7}}}{\alpha_s}\frac{1}{ (Qt)^\frac{4}{7}},\notag\\
&\hat{q}\sim (\alpha_s f_0)^{\frac{4}{7}}\frac{Q^3}{(Qt)^\frac{5}{7}}, \qquad m_D^2 \sim (\alpha_s f_0)^{\frac{3}{7}}\frac{Q^2}{(Qt)^\frac{2}{7}},\qquad T_*\sim\frac{(\alpha_s f_0)^{\frac{1}{7}}}{\alpha_s}\frac{Q}{(Qt)^\frac{3}{7}}\sim pf.
\end{align}
Within $t$, the probability for one hard gluon to double its momentum by merging with another hard gluon is estimated to be $\sim\alpha_s\sqrt{\hat{q}t^2/p} T_*/p\sim1$, which also holds if the BH rate is used. This consistently shows that the hard gluon momentum $p$ does not receive a parametrically larger correction from the inelastic scattering. At $Q t\sim \alpha_s^{-\frac{7}{4}}(\alpha_s f_0)^{-\frac{1}{4}}$, all the above quantities approach their final equilibrium values in eq.~(\ref{eq:quantitieseq}). Therefore, full thermalization is established around this time~\cite{Kurkela:2014tea}.

\subsection{Quantitative results}

\begin{figure}
\includegraphics[width=0.49\textwidth]{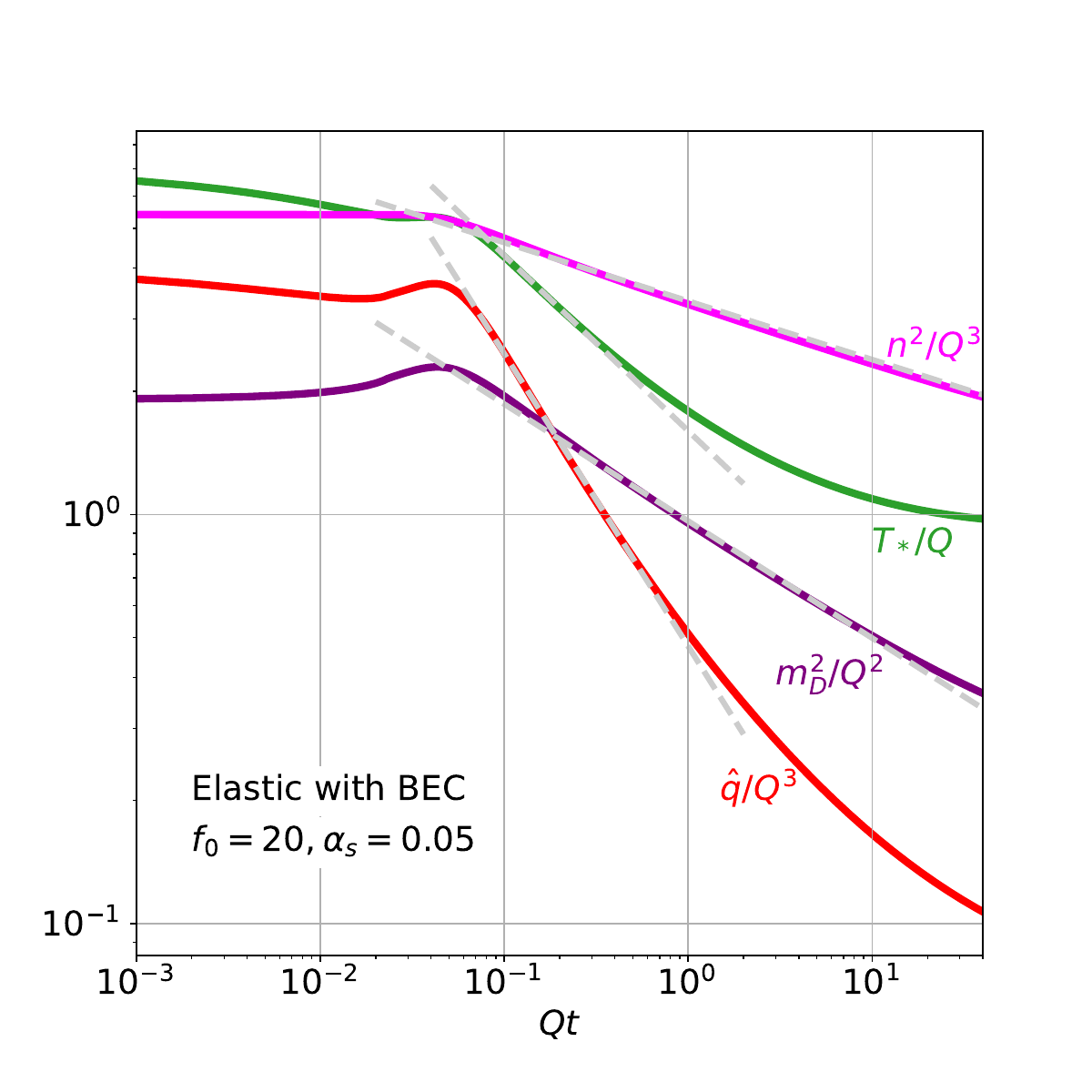}
\includegraphics[width=0.49\textwidth]{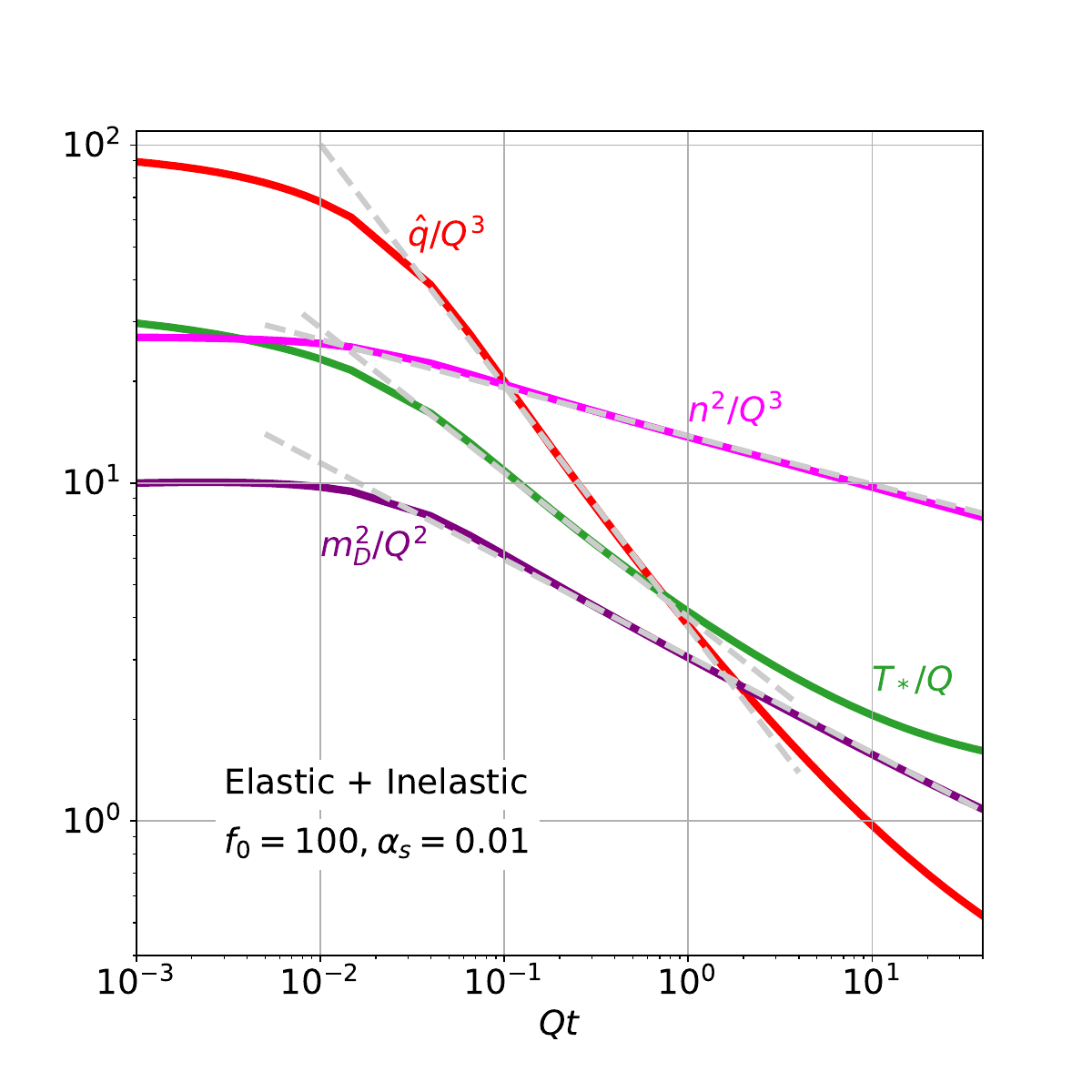}
\caption{The non-thermal scaling solution in over-populated systems. The left panel shows the solution of the BEDA with the elastic kernel only. In this case, soft gluons are taken out as a BEC at $p=0$~\cite{Blaizot:2014jna}. The right panel shows the solution of the BEDA with both the elastic and inelastic kernels in eq.~(\ref{eq:boltz1}). The dash lines represent the power laws in eq.~(\ref{eq:macsOverII}).
}
\label{fig:scaling_over}
\end{figure} 

In the above parametric analysis, momentum broadening due to multiple elastic scatterings plays a major role. The main role of the soft sector during stage 2 is, instead, to effectively reduce the number of gluons as mandated by energy conservation. Different theories such as kinetic theory and classical statistical field theory are different in soft modes, which, however, do not dominate the increment of the hard gluon momentum. Therefore, the scaling laws in eq.~(\ref{eq:macsOverII}) is insensitive to the soft sector. This gives a consistent explanation of the universality of such a scaling solution and its insensitivity to initial conditions as previously discovered both in classical statistical field simulations~\cite{Berges:2012ev, Schlichting:2012es, Kurkela:2012hp} and EKT ~\cite{AbraaoYork:2014hbk, Kurkela:2014tea, Du:2020dvp}.

The two plots of fig.~\ref{fig:scaling_over} show another confirmation of the universality of such a scaling solution. The left panel shows the time evolution of $n, \hat{q}, m_D^2$ and $T_*$ for $f_0=20$ and $ \alpha_s = 0.05$ in BEDA without the inelastic kernel. Our parametric estimates for stage 2, based only on momentum broadening and energy conservation, are also valid in this case (see also ref.~\cite{Blaizot:2011xf}). This plot shows the first numerical confirmation of such scaling laws without the inelastic scattering. As shown in ref.~\cite{Blaizot:2014jna}, the Boltzmann equation demands that a macroscopic number of gluons flow out of the system at $p=0$ at time $t\gtrsim\,t_c\propto (\alpha_s f_0)^{-2}Q^{-1}$. $t_c$ is now understood to be the starting time of such a non-thermal scaling solution. The right panel of this figure shows the scaling behaviors in these quantities for $f_0=100$ and $\alpha_s=0.01$ as a solution of the BEDA including both the elastic kernel and the inelastic kernel with the LPM effect. The two plots also show that the range of $t$ for $f$ to fit the scaling laws becomes wider for larger $f_0$ or, equivalently, smaller $\alpha_s$ when $\alpha_s\,f_0$ is kept fixed. In both cases, the parametric estimates in eqs. (\ref{eq:macsOverI}) and (\ref{eq:macsOverII}) capture the qualitative features of the entire equilibration process.

\section*{Acknowledgements}
The authors would like to thank F. Gelis for reading through the manuscript and informative comments. This work is supported by the European Research Council project ERC-2018-ADG-835105 YoctoLHC; by the Maria de Maetzu excellence program under project CEX2020-001035-M; by the Spanish Research State Agency under project PID2020-119632GB-I00; and by Xunta de Galicia (Centro singular de investigaci\'on de Galicia accreditation 2019-2022), by the European Union ERDF.

\providecommand{\href}[2]{#2}\begingroup\raggedright\endgroup

\begin{thebibliography}{10}

\bibitem{Jalilian-Marian:1996mkd}
J.~Jalilian-Marian, A.~Kovner, L.D.~McLerran and H.~Weigert, \emph{{The
  Intrinsic glue distribution at very small x}},
  \href{https://doi.org/10.1103/PhysRevD.55.5414}{\emph{Phys. Rev. D}
  {\bfseries 55} (1997) 5414}
  [\href{https://arxiv.org/abs/hep-ph/9606337}{{\ttfamily hep-ph/9606337}}].

\bibitem{Kovchegov:1998bi}
Y.V.~Kovchegov and A.H.~Mueller, \emph{{Gluon production in current nucleus and
  nucleon - nucleus collisions in a quasiclassical approximation}},
  \href{https://doi.org/10.1016/S0550-3213(98)00384-8}{\emph{Nucl. Phys. B}
  {\bfseries 529} (1998) 451}
  [\href{https://arxiv.org/abs/hep-ph/9802440}{{\ttfamily hep-ph/9802440}}].

\bibitem{Mueller:1999fp}
A.H.~Mueller, \emph{{Toward equilibration in the early stages after a
  high-energy heavy ion collision}},
  \href{https://doi.org/10.1016/S0550-3213(99)00502-7}{\emph{Nucl. Phys. B}
  {\bfseries 572} (2000) 227}
  [\href{https://arxiv.org/abs/hep-ph/9906322}{{\ttfamily hep-ph/9906322}}].

\bibitem{Schlichting:2019abc}
S.~Schlichting and D.~Teaney, \emph{{The First fm/c of Heavy-Ion Collisions}},
  \href{https://doi.org/10.1146/annurev-nucl-101918-023825}{\emph{Ann. Rev.
  Nucl. Part. Sci.} {\bfseries 69} (2019) 447}
  [\href{https://arxiv.org/abs/1908.02113}{{\ttfamily 1908.02113}}].

\bibitem{Berges:2020fwq}
J.~Berges, M.P.~Heller, A.~Mazeliauskas and R.~Venugopalan, \emph{{QCD
  thermalization: Ab initio approaches and interdisciplinary connections}},
  \href{https://doi.org/10.1103/RevModPhys.93.035003}{\emph{Rev. Mod. Phys.}
  {\bfseries 93} (2021) 035003}
  [\href{https://arxiv.org/abs/2005.12299}{{\ttfamily 2005.12299}}].

\bibitem{Baier:2000sb}
R.~Baier, A.H.~Mueller, D.~Schiff and D.T.~Son, \emph{{'Bottom up'
  thermalization in heavy ion collisions}},
  \href{https://doi.org/10.1016/S0370-2693(01)00191-5}{\emph{Phys. Lett. B}
  {\bfseries 502} (2001) 51}
  [\href{https://arxiv.org/abs/hep-ph/0009237}{{\ttfamily hep-ph/0009237}}].

\bibitem{Berges:2013eia}
J.~Berges, K.~Boguslavski, S.~Schlichting and R.~Venugopalan, \emph{{Turbulent
  thermalization process in heavy-ion collisions at ultrarelativistic
  energies}}, \href{https://doi.org/10.1103/PhysRevD.89.074011}{\emph{Phys.
  Rev. D} {\bfseries 89} (2014) 074011}
  [\href{https://arxiv.org/abs/1303.5650}{{\ttfamily 1303.5650}}].

\bibitem{Epelbaum:2013ekf}
T.~Epelbaum and F.~Gelis, \emph{{Pressure isotropization in high energy heavy
  ion collisions}},
  \href{https://doi.org/10.1103/PhysRevLett.111.232301}{\emph{Phys. Rev. Lett.}
  {\bfseries 111} (2013) 232301}
  [\href{https://arxiv.org/abs/1307.2214}{{\ttfamily 1307.2214}}].

\bibitem{Arnold:2002zm}
P.B.~Arnold, G.D.~Moore and L.G.~Yaffe, \emph{{Effective kinetic theory for
  high temperature gauge theories}},
  \href{https://doi.org/10.1088/1126-6708/2003/01/030}{\emph{JHEP} {\bfseries
  01} (2003) 030} [\href{https://arxiv.org/abs/hep-ph/0209353}{{\ttfamily
  hep-ph/0209353}}].

\bibitem{Kurkela:2015qoa}
A.~Kurkela and Y.~Zhu, \emph{{Isotropization and hydrodynamization in weakly
  coupled heavy-ion collisions}},
  \href{https://doi.org/10.1103/PhysRevLett.115.182301}{\emph{Phys. Rev. Lett.}
  {\bfseries 115} (2015) 182301}
  [\href{https://arxiv.org/abs/1506.06647}{{\ttfamily 1506.06647}}].

\bibitem{Kurkela:2018xxd}
A.~Kurkela and A.~Mazeliauskas, \emph{{Chemical Equilibration in Hadronic
  Collisions}},
  \href{https://doi.org/10.1103/PhysRevLett.122.142301}{\emph{Phys. Rev. Lett.}
  {\bfseries 122} (2019) 142301}
  [\href{https://arxiv.org/abs/1811.03040}{{\ttfamily 1811.03040}}].

\bibitem{Kurkela:2018oqw}
A.~Kurkela and A.~Mazeliauskas, \emph{{Chemical equilibration in weakly coupled
  QCD}}, \href{https://doi.org/10.1103/PhysRevD.99.054018}{\emph{Phys. Rev. D}
  {\bfseries 99} (2019) 054018}
  [\href{https://arxiv.org/abs/1811.03068}{{\ttfamily 1811.03068}}].

\bibitem{Du:2020dvp}
X.~Du and S.~Schlichting, \emph{{Equilibration of weakly coupled QCD plasmas}},
  \href{https://doi.org/10.1103/PhysRevD.104.054011}{\emph{Phys. Rev. D}
  {\bfseries 104} (2021) 054011}
  [\href{https://arxiv.org/abs/2012.09079}{{\ttfamily 2012.09079}}].

\bibitem{Mueller:2002gd}
A.H.~Mueller and D.T.~Son, \emph{{On the Equivalence between the Boltzmann
  equation and classical field theory at large occupation numbers}},
  \href{https://doi.org/10.1016/j.physletb.2003.12.047}{\emph{Phys. Lett. B}
  {\bfseries 582} (2004) 279}
  [\href{https://arxiv.org/abs/hep-ph/0212198}{{\ttfamily hep-ph/0212198}}].

\bibitem{Berges:2013lsa}
J.~Berges, K.~Boguslavski, S.~Schlichting and R.~Venugopalan, \emph{{Basin of
  attraction for turbulent thermalization and the range of validity of
  classical-statistical simulations}},
  \href{https://doi.org/10.1007/JHEP05(2014)054}{\emph{JHEP} {\bfseries 05}
  (2014) 054} [\href{https://arxiv.org/abs/1312.5216}{{\ttfamily 1312.5216}}].

\bibitem{Epelbaum:2014yja}
T.~Epelbaum, F.~Gelis and B.~Wu, \emph{{Nonrenormalizability of the classical
  statistical approximation}},
  \href{https://doi.org/10.1103/PhysRevD.90.065029}{\emph{Phys. Rev. D}
  {\bfseries 90} (2014) 065029}
  [\href{https://arxiv.org/abs/1402.0115}{{\ttfamily 1402.0115}}].

\bibitem{Epelbaum:2014mfa}
T.~Epelbaum, F.~Gelis, N.~Tanji and B.~Wu, \emph{{Properties of the Boltzmann
  equation in the classical approximation}},
  \href{https://doi.org/10.1103/PhysRevD.90.125032}{\emph{Phys. Rev. D}
  {\bfseries 90} (2014) 125032}
  [\href{https://arxiv.org/abs/1409.0701}{{\ttfamily 1409.0701}}].

\bibitem{Mueller:1999pi}
A.H.~Mueller, \emph{{The Boltzmann equation for gluons at early times after a
  heavy ion collision}},
  \href{https://doi.org/10.1016/S0370-2693(00)00084-8}{\emph{Phys. Lett. B}
  {\bfseries 475} (2000) 220}
  [\href{https://arxiv.org/abs/hep-ph/9909388}{{\ttfamily hep-ph/9909388}}].

\bibitem{Hong:2010at}
J.~Hong and D.~Teaney, \emph{{Spectral densities for hot QCD plasmas in a
  leading log approximation}},
  \href{https://doi.org/10.1103/PhysRevC.82.044908}{\emph{Phys. Rev. C}
  {\bfseries 82} (2010) 044908}
  [\href{https://arxiv.org/abs/1003.0699}{{\ttfamily 1003.0699}}].

\bibitem{Blaizot:2013lga}
J.-P.~Blaizot, J.~Liao and L.~McLerran, \emph{{Gluon Transport Equation in the
  Small Angle Approximation and the Onset of Bose-Einstein Condensation}},
  \href{https://doi.org/10.1016/j.nuclphysa.2013.10.010}{\emph{Nucl. Phys. A}
  {\bfseries 920} (2013) 58} [\href{https://arxiv.org/abs/1305.2119}{{\ttfamily
  1305.2119}}].

\bibitem{Blaizot:2014jna}
J.-P.~Blaizot, B.~Wu and L.~Yan, \emph{{Quark production,
  Bose\textendash{}Einstein condensates and thermalization of the
  quark\textendash{}gluon plasma}},
  \href{https://doi.org/10.1016/j.nuclphysa.2014.07.041}{\emph{Nucl. Phys. A}
  {\bfseries 930} (2014) 139}
  [\href{https://arxiv.org/abs/1402.5049}{{\ttfamily 1402.5049}}].

\bibitem{Xu:2014ega}
Z.~Xu, K.~Zhou, P.~Zhuang and C.~Greiner, \emph{{Thermalization of gluons with
  Bose-Einstein condensation}},
  \href{https://doi.org/10.1103/PhysRevLett.114.182301}{\emph{Phys. Rev. Lett.}
  {\bfseries 114} (2015) 182301}
  [\href{https://arxiv.org/abs/1410.5616}{{\ttfamily 1410.5616}}].

\bibitem{Huang:2013lia}
X.-G.~Huang and J.~Liao, \emph{{Glasma Evolution and Bose-Einstein Condensation
  with Elastic and Inelastic Collisions}},
  \href{https://doi.org/10.1103/PhysRevD.91.116012}{\emph{Phys. Rev. D}
  {\bfseries 91} (2015) 116012}
  [\href{https://arxiv.org/abs/1303.7214}{{\ttfamily 1303.7214}}].

\bibitem{Blaizot:2016iir}
J.-P.~Blaizot, J.~Liao and Y.~Mehtar-Tani, \emph{{The thermalization of soft
  modes in non-expanding isotropic quark gluon plasmas}},
  \href{https://doi.org/10.1016/j.nuclphysa.2017.02.003}{\emph{Nucl. Phys. A}
  {\bfseries 961} (2017) 37}
  [\href{https://arxiv.org/abs/1609.02580}{{\ttfamily 1609.02580}}].

\bibitem{Baier:1996kr}
R.~Baier, Y.L.~Dokshitzer, A.H.~Mueller, S.~Peigne and D.~Schiff,
  \emph{{Radiative energy loss of high-energy quarks and gluons in a finite
  volume quark - gluon plasma}},
  \href{https://doi.org/10.1016/S0550-3213(96)00553-6}{\emph{Nucl. Phys. B}
  {\bfseries 483} (1997) 291}
  [\href{https://arxiv.org/abs/hep-ph/9607355}{{\ttfamily hep-ph/9607355}}].

\bibitem{Baier:1998kq}
R.~Baier, Y.L.~Dokshitzer, A.H.~Mueller and D.~Schiff, \emph{{Medium induced
  radiative energy loss: Equivalence between the BDMPS and Zakharov
  formalisms}},
  \href{https://doi.org/10.1016/S0550-3213(98)00546-X}{\emph{Nucl. Phys. B}
  {\bfseries 531} (1998) 403}
  [\href{https://arxiv.org/abs/hep-ph/9804212}{{\ttfamily hep-ph/9804212}}].

\bibitem{Kurkela:2014tea}
A.~Kurkela and E.~Lu, \emph{{Approach to Equilibrium in Weakly Coupled
  Non-Abelian Plasmas}},
  \href{https://doi.org/10.1103/PhysRevLett.113.182301}{\emph{Phys. Rev. Lett.}
  {\bfseries 113} (2014) 182301}
  [\href{https://arxiv.org/abs/1405.6318}{{\ttfamily 1405.6318}}].

\bibitem{Berges:2012ev}
J.~Berges, S.~Schlichting and D.~Sexty, \emph{{Over-populated gauge fields on
  the lattice}}, \href{https://doi.org/10.1103/PhysRevD.86.074006}{\emph{Phys.
  Rev. D} {\bfseries 86} (2012) 074006}
  [\href{https://arxiv.org/abs/1203.4646}{{\ttfamily 1203.4646}}].

\bibitem{Schlichting:2012es}
S.~Schlichting, \emph{{Turbulent thermalization of weakly coupled non-abelian
  plasmas}}, \href{https://doi.org/10.1103/PhysRevD.86.065008}{\emph{Phys. Rev.
  D} {\bfseries 86} (2012) 065008}
  [\href{https://arxiv.org/abs/1207.1450}{{\ttfamily 1207.1450}}].

\bibitem{Kurkela:2012hp}
A.~Kurkela and G.D.~Moore, \emph{{UV Cascade in Classical Yang-Mills Theory}},
  \href{https://doi.org/10.1103/PhysRevD.86.056008}{\emph{Phys. Rev. D}
  {\bfseries 86} (2012) 056008}
  [\href{https://arxiv.org/abs/1207.1663}{{\ttfamily 1207.1663}}].

\bibitem{AbraaoYork:2014hbk}
M.C.~Abraao~York, A.~Kurkela, E.~Lu and G.D.~Moore, \emph{{UV cascade in
  classical Yang-Mills theory via kinetic theory}},
  \href{https://doi.org/10.1103/PhysRevD.89.074036}{\emph{Phys. Rev. D}
  {\bfseries 89} (2014) 074036}
  [\href{https://arxiv.org/abs/1401.3751}{{\ttfamily 1401.3751}}].

\bibitem{Blaizot:2011xf}
J.-P.~Blaizot, F.~Gelis, J.-F.~Liao, L.~McLerran and R.~Venugopalan,
  \emph{{Bose--Einstein Condensation and Thermalization of the Quark Gluon
  Plasma}}, \href{https://doi.org/10.1016/j.nuclphysa.2011.10.005}{\emph{Nucl.
  Phys. A} {\bfseries 873} (2012) 68}
  [\href{https://arxiv.org/abs/1107.5296}{{\ttfamily 1107.5296}}].

\bibitem{Baier:1996sk}
R.~Baier, Y.L.~Dokshitzer, A.H.~Mueller, S.~Peigne and D.~Schiff,
  \emph{{Radiative energy loss and p(T) broadening of high-energy partons in
  nuclei}}, \href{https://doi.org/10.1016/S0550-3213(96)00581-0}{\emph{Nucl.
  Phys. B} {\bfseries 484} (1997) 265}
  [\href{https://arxiv.org/abs/hep-ph/9608322}{{\ttfamily hep-ph/9608322}}].

\bibitem{Arnold:2008zu}
P.B.~Arnold and C.~Dogan, \emph{{QCD Splitting/Joining Functions at Finite
  Temperature in the Deep LPM Regime}},
  \href{https://doi.org/10.1103/PhysRevD.78.065008}{\emph{Phys. Rev. D}
  {\bfseries 78} (2008) 065008}
  [\href{https://arxiv.org/abs/0804.3359}{{\ttfamily 0804.3359}}].

\bibitem{Salgado:2003gb}
C.A.~Salgado and U.A.~Wiedemann, \emph{{Calculating quenching weights}},
  \href{https://doi.org/10.1103/PhysRevD.68.014008}{\emph{Phys. Rev. D}
  {\bfseries 68} (2003) 014008}
  [\href{https://arxiv.org/abs/hep-ph/0302184}{{\ttfamily hep-ph/0302184}}].

\bibitem{Baier:2001yt}
R.~Baier, Y.L.~Dokshitzer, A.H.~Mueller and D.~Schiff, \emph{{Quenching of
  hadron spectra in media}},
  \href{https://doi.org/10.1088/1126-6708/2001/09/033}{\emph{JHEP} {\bfseries
  09} (2001) 033} [\href{https://arxiv.org/abs/hep-ph/0106347}{{\ttfamily
  hep-ph/0106347}}].

\bibitem{Blaizot:2013hx}
J.-P.~Blaizot, E.~Iancu and Y.~Mehtar-Tani, \emph{{Medium-induced QCD cascade:
  democratic branching and wave turbulence}},
  \href{https://doi.org/10.1103/PhysRevLett.111.052001}{\emph{Phys. Rev. Lett.}
  {\bfseries 111} (2013) 052001}
  [\href{https://arxiv.org/abs/1301.6102}{{\ttfamily 1301.6102}}].

\bibitem{Iancu:2015uja}
E.~Iancu and B.~Wu, \emph{{Thermalization of mini-jets in a quark-gluon
  plasma}}, \href{https://doi.org/10.1007/JHEP10(2015)155}{\emph{JHEP}
  {\bfseries 10} (2015) 155}
  [\href{https://arxiv.org/abs/1506.07871}{{\ttfamily 1506.07871}}].

\end{thebibliography}
\end{document}